\documentclass{PoS}
\usepackage{soul}
\usepackage{url}
\usepackage{ae,aecompl}
\usepackage{cleveref}
\usepackage{comment}
\usepackage{pdflscape}
\usepackage[square,numbers,sort&compress]{natbib}

\usepackage{array}

\usepackage{graphicx}	
\usepackage{amsmath}	
\usepackage{amssymb}	
\usepackage{subfig}
\usepackage{caption}

\usepackage{floatrow}
\usepackage{hvfloat}

\title{Testing blazar emission models on the extreme blazar PGC 2402248, newly discovered at very high energies with the MAGIC telescopes}

\ShortTitle{TeV gamma-ray detection of the EHBL PGC 2402248}

\author{\speaker{L. Foffano}$^1$, J. Becerra Gonzalez$^2$, M.~Cerruti$^3$, V. Fallah Ramazani$^4$, E.~Prandini$^5$, F.~Tavecchio$^6$, K.~Asano$^7$, and F.~D'Ammando$^8$ for the MAGIC\footnote{\texttt{https://magic.mpp.mpg.de/}} and Fermi-LAT Collaborations$^9$}

\author{ \\
$^1$University of Padova and Istituto Nazionale di Fisica Nucleare (INFN), Padova, Italy. 
 E-mail: \email{luca.foffano@phd.unipd.it}\\
$^2$ Instituto de Astrofisica de Canarias (IAC) and Universidad de La Laguna, Tenerife, Spain \\
$^3$ Universitat de Barcelona, Barcelona, Spain\\
$^4$ Finnish Centre of Astronomy with ESO (FINCA), University of Turku, Turku, Finland \\
$^5$ Istituto Nazionale di Astrofisica (INAF) and INFN, Padova, Italy \\
$^6$ Istituto Nazionale di Astrofisica (INAF), Merate, Italy \\
$^7$  Institute for Cosmic Ray Research (ICRR), University of Tokyo, Tokyo, Japan \\
$^8$ Istituto Nazionale di Astrofisica (INAF), Bologna, Italy \\
$^9$ For collaboration list see PoS(ICRC2019)1177\\
}

\abstract{
\noindent
Extreme high-energy peaked BL Lac objects (EHBLs) are a new emerging class of blazars. The typical two-hump structured spectral energy distribution (SED) is shifted to higher energies with respect to other more established classes of blazars. Multi-wavelength observations allow us to constrain their synchrotron peak in the medium and hard X-ray bands. Their gamma-ray emission dominates above the GeV gamma-ray band, and in some objects it extends up to several TeV (e.g. 1ES 0229+200). Their hard TeV spectrum is also interesting for the implications on the extragalactic background light indirect measurements, the intergalactic magnetic field estimate, and the possible origin of extragalactic high-energy neutrinos.
Up to now, only a few objects have been studied in the TeV gamma-ray range. In this contribution, we will present the new detection of the EHBL object PGC 2402248, recently discovered in TeV gamma rays with the MAGIC telescopes. The analysis results of a set of multi-wavelength simultaneous observations up to the VHE gamma-ray band provide the broad-band SED of the blazar, which will be used to probe different emission models. Given the extreme characteristics of this blazar, constraints on the physical parameters within the framework of leptonic and hadronic models are derived.
}

\FullConference{36th International Cosmic Ray Conference -ICRC2019-\\
		July 24th - August 1st, 2019\\
		Madison, WI, U.S.A.}

\begin{document}

\section{Introduction}

\noindent
In the blazar family, BL Lac objects are divided into different categories depending on the maximum peak energy of the synchrotron photons they are able to accelerate in their relativistic jets.
The extremely high-energy peaked BL Lac objects \citep[EHBLs,][]{Costamante:2001pu} are blazars able to produce the highest photon energies, and are generally characterized by a synchrotron peak located above~0.3~keV~($\simeq~10^{17}$~Hz). Within the EHBL class, the source 1ES~0229+200 is considered the archetypal object \citep{2001AA...371..512C,aharonian07,tavecchio09}. Thanks to  multi-wavelength (MWL) campaigns performed during the last years, its spectral energy distribution (SED) has been characterized in great detail and shows some of the most important spectral properties of such objects. While the synchrotron peak has been constrained by X-ray observations at about 10 keV \citep{Nustar_EHBLs}, the high-energy hump increases up to about 10 TeV and its peak has not been measured yet. 

The search for the observation of the high-energy hump, peaking in the TeV gamma-ray band, is particularly suitable for the observations with Imaging Air Cherenkov Telescopes (IACTs), like the MAGIC telescopes. 
The characterization of the gamma-ray spectra of EHBLs is crucial in order to unveil the extreme acceleration mechanism taking place within their relativistic jets. However, up to now only few objects classified as EHBLs have been detected in the TeV gamma-ray band. Additionally, within this sample of EHBLs, the objects present significant spectral differences, which prompt further studies for a more appropriate definition of the class \citep{foffano2018}. For this reason, the detection and characterization of new EHBLs in the TeV gamma-ray band - together with the properties extracted from their broad-band SEDs - are fundamental in order to characterize the whole class and to disclose their extreme spectral properties.

In this contribution, the MAGIC Collaboration introduces the first detection in TeV gamma~rays  and the observational results of the source named PGC~2402248.

\section{The EHBL PGC~2402248}

\noindent
The blazar PGC~2402248 was selected thanks to its favourable X-ray and high-energy (HE, E$>$100~MeV) gamma-ray flux that made it a good target for coordinated observations in the very-high-energy (VHE) gamma-ray band.

In the 2WHSP catalog \citep{2whsp}, the source PGC~2402248 (also known as 2WHSP~J073326.7+515354) is reported as having a high synchrotron peak frequency exceeding \mbox{$\nu_{\text{peak,2WHSP}}^{\text{sync}}=10^{17.9}$~Hz}, suggesting an EHBL nature for this object. 

Additionally, this source is associated in the \emph{Fermi}-LAT 3FHL catalog \citep [``Third Catalog of Hard
\emph{Fermi}-LAT Sources'',][]{fermi3FHL} with the source 3FHL~J0733.4+5152 and reports a hard spectral index of \mbox{$\Gamma_{HE} = 1.34 \pm 0.43$}, particularly favourable when extrapolated to TeV gamma-ray energies suitable for Cherenkov telescopes.

\section{The MAGIC telescopes}

\noindent
MAGIC \citep{magicperf_1:2015} is a system of two IACTs able to collect images of the Cherenkov light produced by the electromagnetic cascades initiated when gamma rays interact with the atmosphere. The telescopes site is located at 2200\,m altitude at the Roque de los Muchachos observatory (ORM) on the island of La Palma (Canaries).

\begin{figure}
\centering
\includegraphics[width=0.6\columnwidth]{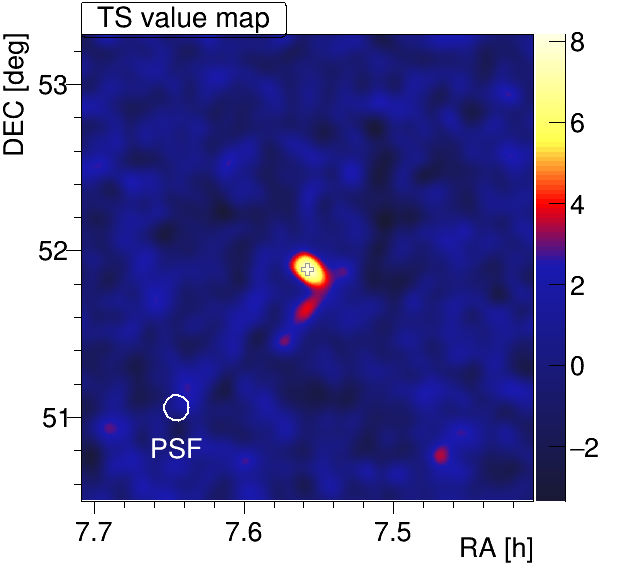}
\caption{TS skymap representing the detection of the source PGC~2402248 with the MAGIC telescopes. The used test statistic is given by Eq. 17 in \cite{LiMa83}, applied on a smoothed and modeled background estimation. Its null hypothesis distribution mostly resembles a Gaussian function, but in general can have a somewhat different shape or width.}
\label{fig:skymap}
\end{figure}

\noindent
The MAGIC telescopes have a field of view of about 3.5\,deg, and the angular resolution reaches 0.1\,deg.
The mirror surface of each telescope has a diameter of 17\,m. This large efective area allows the MAGIC telescopes to lower the energy threshold to about 50 GeV when observing at low zenith angles and in dark nights. Point-like sources, when assuming Crab Nebula-like spectrum, can be detected with an integral sensitivity up to $0.66\pm0.03$\% of the Crab Nebula flux in 50~hours of observation. More details about the instrument's performance and details are described in \cite{MAGICsens} and references therein.

\section{Results and conclusions}

\noindent
The MAGIC telescopes observed the source PGC~2402248 obtaining about 23~hours of good quality data. The observations were performed during 25 nights from January 23rd to April 19th, 2018 (MJD 58141-58227), with zenith angle range between 23$^\circ$ and 40$^\circ$ under dark sky conditions, and lead to the first detection of the source in the TeV gamma-ray band.
In \Cref{fig:skymap}, we report the Test Statistics (TS) skymap that confirms the presence of an excess in the direction of PGC~2402248.

During the MAGIC observational campaign, MWL coordinated observations have been performed and simultaneous data have been obtained with several facilities such as the \emph{Neil Gehrels Swift} satellite, the  Kungliga Vetenskapsakademien Academy (KVA) telescope, and the \emph{Fermi}-LAT instrument. The collection of such MWL data allowed us to build a simultaneous broad-band SED of the source. A new estimation of the synchrotron peak frequency of PGC 2402248 is reported in this work as $\nu_{\text{peak}}^{\text{sync}} \simeq {10^{17.8 \pm 0.3}}$ Hz, being compatible with the estimation reported in the 2WHSP catalog of \mbox{$\nu_{\text{peak,2WHSP}}^{\text{sync}}=10^{17.9}$~Hz} and confirming the classification of PGC 2402248 as an EHBL.

Several theoretical models have been applied to the SED in order to test different acceleration mechanisms within the blazar jet, discussing the possible leptonic and hadronic origin of the processes taking place in the emission region.
The standard Synchrotron-Self-Compton model \cite[e.g.][]{Maraschi:SSC,Tavecchio:SSC} reports extreme parameter values such as a low magnetic field, a large Doppler factor, and a high minimum energy of the electrons.
Such tension in the parameter space can be partially recovered by using more complex theoretical models in which a two-zone structure or hadronic contributions to the emission process are considered.

More details on the results will be published soon in a dedicate paper by the MAGIC Collaboration.


\section{Acknowledgements}

\noindent
We would like to thank the Instituto de Astrof\'{\i}sica de Canarias for the excellent working conditions at the Observatorio del Roque de los Muchachos in La Palma. The financial support of the German BMBF and MPG, the Italian INFN and INAF, the Swiss National Fund SNF, the ERDF under the Spanish MINECO (FPA2015-69818-P, FPA2012-36668, FPA2015-68378-P, FPA2015-69210-C6-2-R, FPA2015-69210-C6-4-R, FPA2015-69210-C6-6-R, AYA2015-71042-P, AYA2016-76012-C3-1-P, ESP2015-71662-C2-2-P, FPA2017-90566-REDC), the Indian Department of Atomic Energy, the Japanese JSPS and MEXT, the Bulgarian Ministry of Education and Science, National RI Roadmap Project DO1-153/28.08.2018 and the Academy of Finland grant nr. 320045 is gratefully acknowledged. This work was also supported by the Spanish Centro de Excelencia ``Severo Ochoa'' SEV-2016-0588 and SEV-2015-0548, and Unidad de Excelencia ``Mar\'{\i}a de Maeztu'' MDM-2014-0369, by the Croatian Science Foundation (HrZZ) Project IP-2016-06-9782 and the University of Rijeka Project 13.12.1.3.02, by the DFG Collaborative Research Centers SFB823/C4 and SFB876/C3, the Polish National Research Centre grant UMO-2016/22/M/ST9/00382 and by the Brazilian MCTIC, CNPq and FAPERJ.

The \textit{Fermi}-LAT Collaboration acknowledges generous ongoing support from a number of agencies and institutes that have supported both the development and the operation of the LAT as well as scientific data analysis. These include the National Aeronautics and Space Administration and the Department of Energy in the United States, the Commissariat \`a l'Energie Atomique and the Centre National de la Recherche Scientifique / Institut National de Physique Nucl\'eaire et de Physique des Particules in France, the Agenzia Spaziale Italiana and the Istituto Nazionale di Fisica Nucleare in Italy, the Ministry of Education, Culture, Sports, Science and Technology (MEXT), High Energy Accelerator Research Organization (KEK) and Japan Aerospace Exploration Agency (JAXA) in Japan, and the K.~A.~Wallenberg Foundation, the Swedish Research Council and the Swedish National Space Board in Sweden.
 
Additional support for science analysis during the operations phase is gratefully acknowledged from the Istituto Nazionale di Astrofisica in Italy and the Centre National d'\'Etudes Spatiales in France. This work performed in part under DOE Contract DE-AC02-76SF00515.

We acknowledge the use of public data from the \emph{Swift} data archive. This publication makes use of data products from the Wide-field Infrared Survey Explorer, which is a joint project of the University of California, Los Angeles, and the Jet Propulsion Laboratory/California Institute of Technology, funded by the National Aeronautics and Space Administration. We also acknowledge the use of the Space Science Data Base (SSDC).

J. Becerra Gonz\'alez acknowledges the support of the Viera y Clavijo program funded by ACIISI and ULL. M. Cerruti has received financial support through the Postdoctoral Junior Leader Fellowship Programme from la Caixa Banking Foundation, grant  n. LCF/BQ/LI18/11630012

\bibliographystyle{mnras}
\bibliography{bibpaper}

\end{document}